\documentclass[a4paper,fleqn,usenatbib]{mn2e}

\usepackage[T1]{fontenc}
\usepackage{ae,aecompl}

\usepackage{graphicx}	
\usepackage{amsmath}	
\usepackage{amssymb}	
\usepackage{color}

\newcommand{\Msun}{\mbox{\,M$_\odot$}}

\newcommand{\Lsun}{\mbox{\,L$_\odot$}}
\newcommand{\vunit}{\mbox{\,km\,s$^{-1}$}}
\newcommand{\mic}{\mbox{$\,\mu$m}}

\newcommand{\fion}[2]{[{#1}\,{\sc {#2}}]}

\newcommand{\au}{\mbox{\,\sc au}}

\newcommand{\ck}{\mbox{CK~Vul}}

\newcommand{\sirtf}{\mbox{\it Spitzer}}

\title[\ck: dwarf merger in the 1670s]{ ALMA reveals the aftermath of
  a white dwarf--brown dwarf merger in \ck{peculae}}
\author[S. P. S. Eyres et al.]{S. P. S. Eyres$^{1,2}$,
A. Evans$^3$,
A. Zijlstra$^4$,
A. Avison$^4$,
R. D. Gehrz$^5$,
M. Hajduk$^6$,\newauthor
S. Starrfield$^7$,
S. Mohamed$^8$,
C. E. Woodward$^5$,
R. M. Wagner$^{9,10}$
\\
$^{1}$Faculty of Computing, Engineering \& Science, University of South Wales, Pontypridd CF37 1DL, UK\\
$^{2}$Jeremiah Horrocks Institute, University of Central Lancashire, Preston PR1 2HE, UK\\
$^{3}$Astrophysics Group, Keele University, Keele, Staffordshire ST5 5BG, UK\\
$^{4}$Jodrell Bank Centre for Astrophysics, School of Physics and Astronomy, University of Manchester, Manchester, M13 9PL, UK\\
$^{5}$Minnesota Institute for Astrophysics, School of Physics \& Astronomy,
116 Church Street SE, University of Minnesota, \\Minneapolis, MN 55455, USA\\
$^{6}$Space Radio-Diagnostics Research Centre, University of Warmia and Mazury,
Prawochenskiego Str. 9, 10-720 Olsztyn, Poland\\
$^{7}$School of Earth and Space Exploration, Arizona State University, Box 871404, Tempe, AZ
85287-1404, USA\\
$^{8}$South African Astronomical Observatory, PO Box 9, Observatory 7935, Cape Town, Western Cape, South Africa\\
$^{9}$Department of Astronomy, The Ohio State University, 140 West 18th Avenue, Columbus, OH 43210, USA\\
$^{10}$LBT Observatory, University of Arizona, Tucson, AZ 85721-0065, USA
}

\date{Accepted 14th September 2018. Received YYY; in original form
  14th April 2018}

\pubyear{2018}

\begin{document}
\label{firstpage}
\pagerange{\pageref{firstpage}--\pageref{lastpage}}
\maketitle

\begin{abstract} 
  We present Atacama Large Millimeter--Submillimeter Array (ALMA)
  observations of \ck{peculae} which is identified with ``Nova
  Vulpeculae~1670''. They trace obscuring dust in the inner regions of
  the associated nebulosity. The dust forms {two cocoons}, each
  extending $\sim5\arcsec$ north and south of the presumed location of
  the central star. {Brighter emission is in a more} compact
  east--west structure ($2\arcsec\times1\arcsec$) where the {cocoons}
  intersect. We detect line emission in NH$_2$CHO, CN, {four} organic
  molecules and C$^{17}$O. {CN} lines trace bubbles within the dusty
  cocoons; {CH$_3$OH} a north--south S--shaped jet; and {other
    molecules} a central cloud with a structure aligned with the
  {innermost dust structure}. The major axis of the overall dust and
  gas bubble structure has a projected inclination of $\sim24^\circ$
  with respect to a $71\arcsec$ extended ``hourglass'' nebulosity,
  previously seen in H$\alpha$. Three {cocoon} limbs align with dark
  lanes in the inner regions of the same H$\alpha$ images. The central
  {$2\arcsec\times1\arcsec$} dust is resolved into a structure
  consistent with a warped dusty disc. The velocity structure of the
  jets indicates an origin at the centre of this disc and precession
  with an unknown period. Deceleration regions at both the northern
  and southern tips of {the} jets are roughly coincident with
  additional diffuse dust emission over regions approximately
  $2\arcsec$ across.  These structures are consistent with a bipolar
  outflow expanding into surrounding high density material.  We
  suggest that {a white dwarf and brown dwarf merged between 1670 and
    1672}, with the observed structures {and extraordinary isotopic
    abundances} generated as a result.
\end{abstract}





\section{Introduction}
\label{sec:intro}


Nova~Vul~1670 was discovered at 3rd magnitude by P\`ere Dom Anthelme
on 20 June 1670, and is described in contemporary accounts as {\it
  Nova sub Capite Cygni} \citep*[see for example Figure~1
of][]{shara}. Using H$\alpha$ imaging, \cite{SM} claimed to have
recovered the remnant of the 1670 event\footnote{We refer to the 1670
  event as ``Nova Vul~1670'' and to the object observed post--1980 as
  \ck.}  and had it proven to have been a classical nova, the object
might have provided a means of investigating the inter--outburst
behaviour of this class.  In a detailed study, \cite{shara}
reconstructed the original observations of Nova~Vul~1670 in the form
of a light curve calibrated to the modern magnitude scale, showing two
peaks brighter than 3$^{\rm rd}$ magnitude in 1670 and 1671.
However \cite{shara} recognised that there are serious shortcomings
with this interpretation. For example the light curve is extremely
un--nova--like, and the expansion velocity implied by the current size
of the remnant is implausibly slow, even for a very slow classical
nova. They and others have considered several alternative
interpretations, which have (over time) included a Herbig--Haro
object, a late thermal pulse, a ``diffusion--induced nova'', and a
stellar merger \citep[see][for
discussion]{smorgasbord,kaminski-n,kaminski-a}.


\cite{hajduk-ck} found a radio source at the position of \ck. They
interpreted this as optically thin free--free emission, implying that
the central source must be hot enough to ionise the nebula; on the
assumption that the radio emission is free--free, they estimate a mass
of $4\times10^{-7}$\Msun\ for the ionised gas.  They also discovered a
faint extended bipolar ($\sim71\arcsec$) H$\alpha$ nebula, presenting
evidence that it is expanding, with an origin -- centred on the radio
source -- consistent with a 1670 ejection event.  They speculated that
the radio source could be associated with a circumbinary disc, similar
to those seen in some binary post--AGB stars.

\cite{smorgasbord} presented {\em Spitzer}~IRS spectroscopy in the
range 5.8--40.0~$\mu$m.  They found emission from HCN, H$_2$,
 \fion{O}{iv}, \fion{Si}{ii} and \fion{Si}{iii}, together with
Unidentified Infrared (UIR) emission -- consistent with dust formed in
a carbon--rich environment.  Excitation of the UIR features requires a
source of ultraviolet radiation; the central wavelength of the
`7.7~$\mu$m' and `11.2~$\mu$m' UIR features imply an effective
temperature of at least $\sim$14\,000~K.

From the arguments in \cite{smorgasbord} regarding the UIR features
and the presence of ionised species such as \fion{O}{iv}, the
H$\alpha$ emission, and the free--free radio source \citep{hajduk-ck},
there is a considerable amount of evidence to suggest that there is a
source of ultraviolet radiation at the heart of \ck.

\cite{kaminski-n} observed \ck\ with the Atacama Pathfinder Experiment
\citep[APEX;][]{apex}, and found a rich spectrum of diatomic and
triatomic molecules; they used isotopologues to estimate the
$^{12}$C/$^{13}$C, $^{14}$N/$^{15}$N and $^{16}$O/$^{18}$O isotopic
ratios, all of which are grossly non--solar (and non--nova).  These
isotope ratios suggest non--equilibrium CNO burning, possibly at an
elevated temperature.  \citeauthor{kaminski-n} also observed \ck\ with
the Submillimeter Array \citep[SMA;][]{sma}, finding both a jet
structure in CO(3--2) emission with an extent of $13\arcsec$ north--south,
and dust emission arising from the same region as the radio emission
detected by \cite{hajduk-ck}. The dust source has structure
$\sim3.7\arcsec\times\sim1.0\arcsec$ at PA $33^\circ$, but there is other
structure as well. They suggested that the dust emission arises in a
torus/disc and a pair of jets.

Further observations with APEX and IRAM \citep{kaminski-a} extend the
spectroscopy over a range from $\sim$70 to $\sim$900~GHz, with spatial
resolution at frequency $\nu$ of $8073\arcsec/\nu$ (APEX) and $3393\arcsec/\nu$
(IRAM), or between $\sim9\arcsec$ and $\sim48\arcsec$ (900 down to 70~GHz). They
identify emission from 27 molecules, comprising the elements H, C, N,
O, F, Al, Si, P, and S.

Here we present high spatial resolution imaging of the dust emission
from the inner $\sim24\arcsec$ of the system, taken with the Atacama
Large Millimeter--Submillimeter Array \citep[ALMA;][]{alma}. We
examine the dust emission as well as line emission from gas in a range
of structures, extending north and south of the central source
approximately {$5\arcsec$ in each direction}.

\section{Observations}
\label{sec:observations}


Observations were made of \ck\ with ALMA at Band~6, in four spectral
windows 1.875~GHz in width and centred on 224~GHz, 226~GHz, 240~GHz
and 242~GHz. These are referred to as SPW0, SPW1, SPW2 and SPW3
respectively for brevity. Data were collected in 128~channels in each
window, giving a modest velocity resolution slightly better than
21~km~s$^{-1}$. Channels 0--7 and 119--127 (SPW0 and SPW1), and
channels 0--8 and 120--127 (SPW2 and SPW3), sit outside the system
bandpass.


Our selected sensitivity and spatial resolution targets were met with
two observations: 10~min 8~sec in array configuration C40--3; and
23~min 18~sec in an intermediate array of approximately the same
antenna distribution as array configuration C40--6. The latter data
are referred to as the C40--6 data for brevity, although in practice
the longest baseline exceeded that of the standard configuration.


The C40--3 data were taken starting at 2017~April~24 11:47:53~UT, using
J2148+0657 as flux calibrator \citep[see][for a discussion of ALMA
calibrators]{vanK}, J2025+3343 as bandpass calibrator and J1935+2031
as phase calibrator.  The array incorporated 39 antennae, with a
maximum baseline of 460~m and minimum baseline of 15.1~m.


The C40--6 data were taken starting at 2017~July~22 03:26:07~UT, using
J1751+0939 as flux and bandpass calibrators, and J1952+2526 as phase
calibrator.  The array incorporated 42 antennae, with a maximum
baseline of 3.7~km and minimum baseline of 16.7~m.


The data were recovered from the ALMA archive and calibrated using the
standard scripts provided by the pipeline. The two sets of calibrated
data were then combined into a single $uv$ measurement set using {\sc
  casa} task concat\citep{mcmullin2007}, with default settings except
copypointing=False as we are not mosaicing.


By inspection of the calibrated amplitude versus frequency in each
spectral window, as provided by the calibration pipeline, we
identified the following channels as containing continuum emission:
SPW0 25 to 118; SPW1 50 to 75 and 90 to 118; SPW2 all; and SPW3 9 to
35 and 95 to 119. 


The continuum channels were used to derive a map of the dust
emission, using {\sc casa} task tclean. Briggs weighting was applied
with robust=2. The resultant image has a beam of 280~milliarcsec by
250~milliarcsec, with the major axis at position angle $-177^\circ$. 

\begin{figure*}
\includegraphics[width=\textwidth]{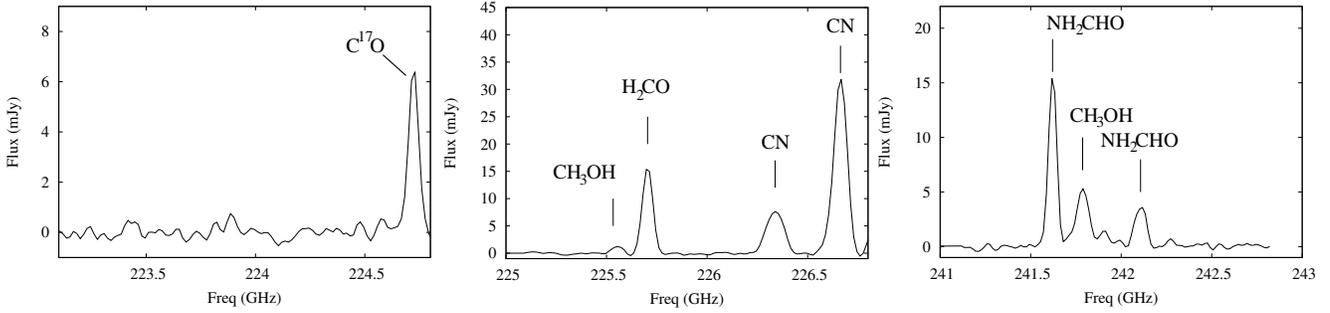}
\caption{Integrated spectra indicating the eight lines listed
  in Table~\ref{IDs}. Note the additional line rising at the edge of the band just above the strong CN line at 226.66~GHz; this line contributes to the high
  velocity emission depicted in Fig~\ref{maps1}(d). 
    \label{lines}}
\end{figure*}

These channel ranges were then used to remove the continuum data in
the $uv$ plane, using uvcontsub. The continuum--subtracted data in
each spectral window were then imaged in spectral cube mode to allow
examination of the emission.  The spatial resolution varies slightly
across the channels in the four spectral windows, but is close enough
to the continuum resolution to be considered identical for the
purposes of comparison.  We detect eight distinct lines showing
extended emission.  We have identified these lines as outlined in Table~\ref{IDs}{, and show an integrated spectrum in 
Fig~\ref{lines}}.

\begin{table}
\caption{Line identifications. \label{IDs}}
 \begin{tabular}{cccc}
Line &            Observed central  & Rest                &  {Line}     \\
         &            frequency (GHz)   & frequency$^*$ (GHz) & identification  \\ \hline
 1       &     $224.273\pm0.001$        & 224.695             & C$^{17}$O      \\
 2       &     $225.553\pm0.003$        & 225.524             & CH$_3$OH       \\
 3       &     $225.704\pm0.001$        & 225.675             & H$_2$CO        \\
 4       &     $226.338\pm0.001$        &  226.309            & CN            \\
 5       &     $226.663\pm0.001$        & 226.634             & CN            \\
 6       &     $241.621\pm0.001$        & 241.590             & NH$_2$CHO        \\
 7       &     $241.787\pm0.001$        &  241.756            & CH$_3$OH      \\
 8       &     $242.107\pm0.001$        & 242.076             & NH$_2$CHO      \\  \hline
 \multicolumn{4}{l}{$^*$Assuming $v_{\rm LSR}=-38$\vunit\ calculated for 2017~April~24.} \\
 \end{tabular}
\end{table}

A spectral cube was extracted for each line across all channels
showing emission from that line. The frequency of the fitted peak for
the corresponding line (given in Table~\ref{IDs}) was then assigned
the $v_{\rm LSR}=-38$\vunit, obtained using the Starlink rv package
\citep{wallace,currie}. The frequency of each line differs from
  the $v_{\rm LSR}$--adjusted frequency by less than the spectral
  resolution, except 
  \textcolor{black}{for the line at observed 
  frequency\footnote{\textcolor{black}{All frequencies in the discussion that follows are the observed frequencies in Table~\ref{lines}.}} 225.553~GHz (CH$_3$OH,
  but see below for a discussion of the}
  emission distribution for this line). This allows us to understand
the velocity structure in comparison to the local standard of
rest. For the cube containing 
\textcolor{black}{the 226.663~GHz line}  (see Table~\ref{IDs}), structure
at the edge of the band is associated with an adjacent line centred
beyond the edge of the spectral window at 226.85~GHz{, as can be
  seen at the edge of the middle panel of Fig~\ref{lines}}. The rest
frequency for that feature differs from the value assigned to that
cube, and is unknown as the centre is not in the band.


We defer modelling of the dynamics of these emission lines to a
future, more detailed, paper; for our present purposes we confine the
discussion to a comparison of the distribution of selected lines with
the dust emission.

\section{Results}
\label{sec:results}


\cite{kaminski-n} associate the continuum emission around 230~GHz with
dust, based on the spectral energy distribution (SED). They report a
central distribution extended north--south and east--west. Taken
together this allows us to associate the continuum emission in
Fig.~\ref{maps}(a) with the dust distribution, the central peak being
aligned with the point of origin for the expansion discussed by
\cite{hajduk-ck}. The axis of the north--south extension is
consistent with that of the CO jet identified by \cite{kaminski-n}.

\subsection{The dust}
\label{dust}

Determination of the dust mass (see Section~\ref{dust-mass} below)
requires the distance to \ck. For consistency with most of the work on
this object, we adopt a value of $D=700$~pc throughout
\citep{hajduk-ck2,kaminski-n}. We also assume a continuum observing
wavelength of 1249\mic, which at 240.2~GHz sits in SWP2 and is
representative for the continuum.

\subsubsection{The dust distribution}
\begin{figure*}
\includegraphics[width=\textwidth]{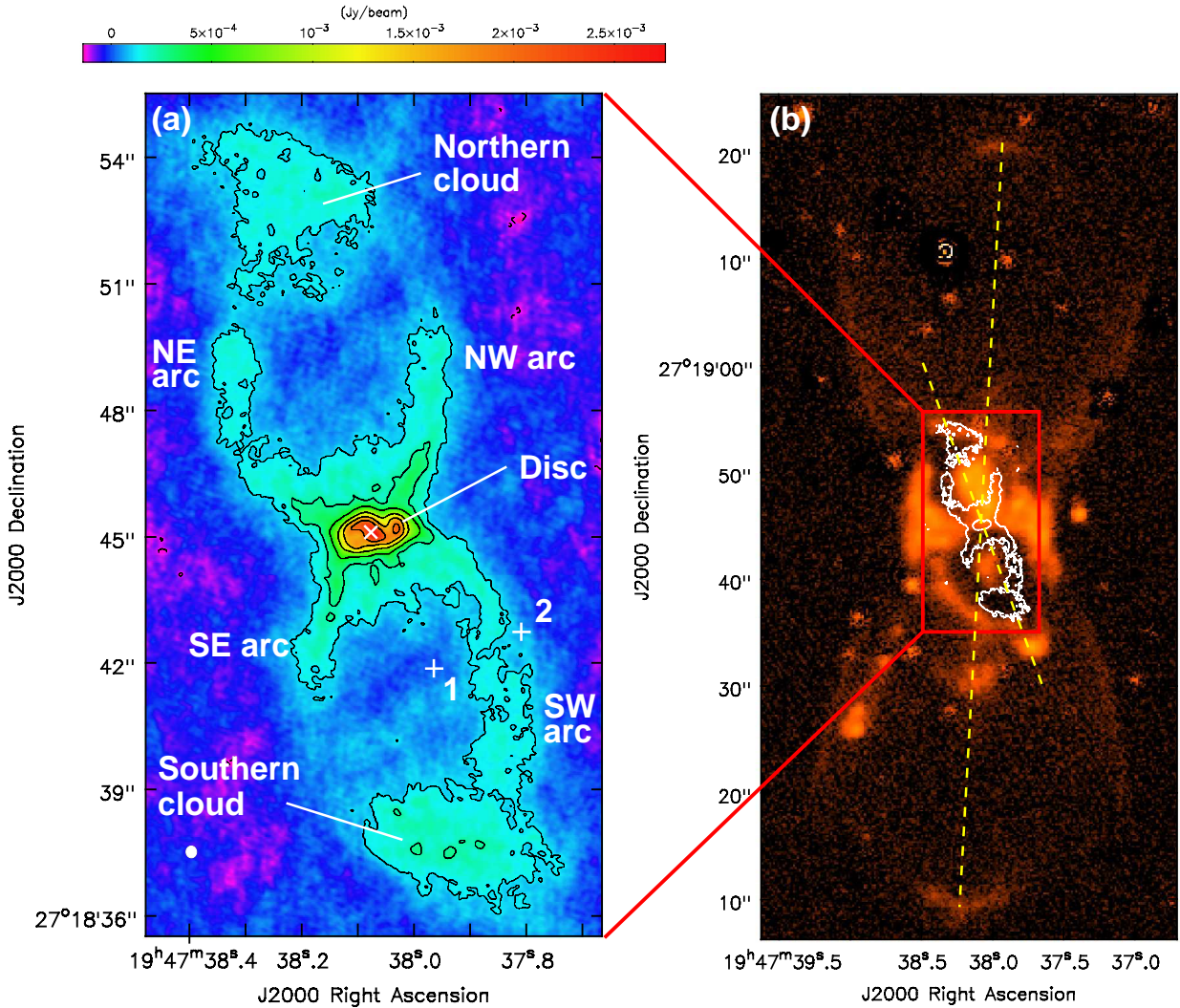}
\caption{Dust emission from \ck\ as observed by ALMA. (a)
    continuum emission associated with dust, contours at $-0.12$,
    0.12, 0.24, 0.48, 0.96, 1.44 and 1.92~mJy~beam$^{-1}$.  The colour
    scale descends to 3$\sigma$ and shows the emission between the
    arcs to north and south. Spatial resolution is represented by the
    small white circle in the bottom left of the image. The peak of
the free--free emission reported by Hajduk et ql. (2007) is marked
with an X, the location of the two variable stars discussed in
    that paper and in Section~\ref{dust} are marked with crosses,
    labelled 1 and 2. (b) Dust emission (contours at 0.12 and
    0.48~mJy~beam$^{-1}$ for reference) in relation to the extended
H$\alpha$ emission (Hajduk et al. 2007). The white ring at 19:47:38.3
27:19:10 is a saturated star in the H$\alpha$ image. Red solid
    lines are included to demonstrate the relative scales of the two
    imaged regions. Dashed yellow lines indicate the major axes of the
    dust and H$\alpha$ emission referred to in the text.
    \label{maps}}
\end{figure*}

There are three distinct components to the dust distribution (see
Fig~\ref{maps}{(a)}):
\begin{enumerate}
\item Four ``arcs'' of emission that extend $4\arcsec$ to $6\arcsec$, two each
  to the north and south, and which in three cases align well with
  dark lanes in the H$\alpha$ emission reported by \cite{hajduk-ck}
  (see Fig.~\ref{maps}{(b)}).
\item Some $6\arcsec$ to both north and south, there are faint, irregular
  but elongated regions (``clouds'') approximately
  $3\arcsec\times4\arcsec$.  The ``major axes'' of these features are oriented
  NE--SW, and are approximately orthogonal to the {NE and SW
  arcs at the point the arcs appear to intersect with the clouds.}
\item The inner $\sim2\arcsec$ is extended east--west with a substructure
  that includes north--south extension around the peak, suggestive of
  a warped disc. {We refer to this as the compact region, to
    differentiate from the diffuse emission in the arcs and clouds.}
  The free--free radio source reported by \cite{hajduk-ck} coincides
  with the brightest part of this feature {within the errors, and
    the peak position of this is marked with an X in
    Fig.~\ref{maps}(a)}.
\end{enumerate}

The dust emission has approximate point symmetry about the position of
the radio source, and hence the putative stellar remnant.  The central
component resembles a warped disc, commonly seen in protoplanetary
discs, where they may be the result of irradiation
\citep[e.g.][]{nixon} or the presence of massive planetary bodies
\citep[e.g.][]{walsh}.

The major axes of diffuse dust emission {(the arcs and clouds)}
and those of the H$\alpha$ hourglass reported by \cite{hajduk-ck} are
misaligned (see Fig.~\ref{maps}{(b), in which we mark these two
  axes as dashed lines}): the major axis of the dust lobes has
position angle $\sim{13}^\circ$, whereas that of the tip of the
H$\alpha$ emission is $\sim{349}^\circ$. If these features are due to
outflows of essentially the same nature, but ejected at different
times and hence at different points in the evolution of the central
object, there has been some precession in the intervening time,
amounting to $\sim24^\circ$ projected on to the sky.  This may be
related to the warped disc, noted above. This is consistent with the
interpretation of a similar misalignment as precession, by
\cite{kaminski-n}

The distribution of the dust in space, rather than that projected on
the sky, can be characterised from the observed variation across the
structure. If it is arranged in four {narrow} ``filaments'', we
would expect to see essentially no emission in between each pair of
arcs. On the other hand if the dust is distributed in hourglass--like
{``cocoons''}, then optically thin emission from the arcs (of
width $\Delta R$) should be a factor
$\sim\sqrt{2R/\Delta{R}+1}\sim2.3$ brighter than from between the arcs
(separated by distance $R$). The latter is close to what is observed,
suggesting the observed arcs are projections of {these dust
  cocooons.

The two variable stars discussed by \cite{hajduk-ck}, are
  marked on Fig.~\ref{maps}(a), at the positions in
  Table~\ref{stars}}.
\begin{table}
  \caption{Positions (J2000) of two variable stars from Hajduk, van Hoof \& Zijlstra (2013) and marked in Fig.~\ref{maps}.
\label{stars}}
\begin{center}
 \begin{tabular}{lcc}
Star & RA & Dec\\
\hline
1 & 19:47:37.964 & +27:18:41.86\\
2 & 19:47:37.811 & +27:18:42.74\\
\hline
\end{tabular}
\end{center}
\end{table}
{The variability was
  attributed to a varying dust distribution in the foreground, and the
  proposed dust was associated with \ck. They further proposed that
  the lithium absorption in both stars arises from the ejecta from
  Nova~Vul~1670. The ALMA observations place the eastern star (1) behind the central
  cavity of the dust cocoon, while the western star (2) is at the western
  edge of the cocoon. Thus the interpretation by \cite{hajduk-ck} would seem consistent
  with the resolved dust structure. In R band, \cite{hajduk-ck} found
  star 1 to have brightened by around 1.5~mag between 1983 and 2010,
  while star 2 declined by around 2~mag between 1991 and 2010. This
  would indicate a reduction in the dust along the line of sight to
  star 1, and an increase along the line of sight to star 2. This
  would be consistent with the southern jet currently depositing dust
  along the line of sight towards star 2, while the dust in front of star 1 is dispersing now
  that the jet has precessed to a different position.

}

\subsubsection{Dust mass and the gas--to--dust ratio} 
\label{dust-mass}

Due to the irregular distribution of the dust emission, we have
estimated the flux density using a ``pixel counting'' method. For the
inner region shown in Fig.~\ref{maps}(a) we added up the total
flux included within the 0.48~mJy~beam$^{-1}$ contour; this included
2226 pixels and gave a total flux density of 31.31$\pm$0.04~mJy. For
the entire dust emission, we added up the total flux included within
the 0.12~mJy~beam$^{-1}$ contour, a region containing 61903 pixels,
finding a value of 132.84$\pm$0.04~mJy. Estimates of uncertainty were
made by calculating the rms brightness in regions containing no dust
emission. We note this is higher than the value of
  75.0$\pm$0.4~mJy measured by \cite{kaminski-n}, but we image the
  dust over a larger area. From the difference between the total dust
emission and that in the inner region, we estimate that the emission
from the diffuse regions alone amounts to 101.53$\pm$0.06~mJy.

\cite{kaminski-n} have a dual grey body model for the \ck\ mm/sub--mm SED,
with dust temperatures $T=15$~K and $T=49$~K. The IR--mm SED in
these papers suggests that the ALMA emission lies on the
Rayleigh--Jeans tail for both temperatures. Given the abundance of
hydrocarbons in the environment of \ck\ \citep{smorgasbord,
  kaminski-n}, we use the mass absorption coefficient for amorphous
carbon (``AC'') from \cite{menella}, who give a value
66~cm$^2$~g$^{-1}$ at 1100\mic. Thus
\begin{equation}
\frac{M_{\rm dust}}{\Msun} \simeq 3.18\times10^{-5} \left ( \frac{D}{\rm kpc} \right )^2
                                                  \left ( \frac{f_\nu}{\rm mJy} \right )
                                                  \left ( \frac{T}{\rm K} \right )^{-1}
                                                  \left ( \frac{\lambda}{1100\mic} \right )^{2.9} \label{Mdust} 
\end{equation}
for the dust mass. The SED in \citeauthor{kaminski-n} shows the 15~K
emission dominates at 230~GHz.  Using this as the temperature of the
dust in Fig.~\ref{maps}(a), we find a total mass of $M_{\rm dust} \sim
2.04\times10^{-4}$\Msun. Of this $\sim 1.56\times10^{-4}$\Msun\ is in
the diffuse extended emission, and $\sim 4.81\times10^{-5}$\Msun\ is
in the central disc.

\cite{kaminski-n} quote a gas mass of $\sim1$\Msun, which results in a
gas--to--dust ratio $\sim10^4$ by mass, a value that is 10--100 times
that seen in similar nebulae \cite[e.g.][]{sarkar,otsuka,walsh-j}.
Assuming the CO column density in \citeauthor{kaminski-n}, and a
CO/H$_2$ ratio of $\sim10^{-3}$ (appropriate for carbon--rich AGB
winds), we recalculated a H$_2$ mass of
$\sim9.8\times10^{-3}$\Msun. {We note that while most of the species
identified by \cite{kaminski-a} are carbon bearing, there are also
numerous oxygen bearing molecules in their list. If the nebula is
considered oxygen--rich, this will give a value
$\sim3.9\times10^{-2}$\Msun.}

{These are} likely to be a lower limit on the hydrogen mass as
much of this may be in atomic and ionic form.  But if we assume
$M_{\rm gas} \sim 9.8\times10^{-3}$ {to} 
${3.9}\times10^{-2}$\Msun, and using the total dust mass estimate from
the ALMA observation, the gas--to--dust ratio is $\sim50$ {to 200,
  which seems more reasonable. Thus calculating the H$_2$ ratio in a
  manner consistent with the observed nature of the nebula leads to a
  mass estimate that is also consistent with the dust mass we
  determine. \cite{kaminski-a} justifiably describe \ck\ as having an
  ``extraordinary isotopic composition'', which may imply any adopted
  typical CO/H$_2$ ratio could be misleading, but there is no basis
  for adopting the gas--to--gas ratio of $\sim10^4$ required for a
  mass of 1~\Msun. Thus a value of 0.01 to 0.1\Msun would seem most
  plausible given the other derived mass estimates.}

\subsection{The lines}

\begin{figure*}
\includegraphics[width=0.45\textwidth]{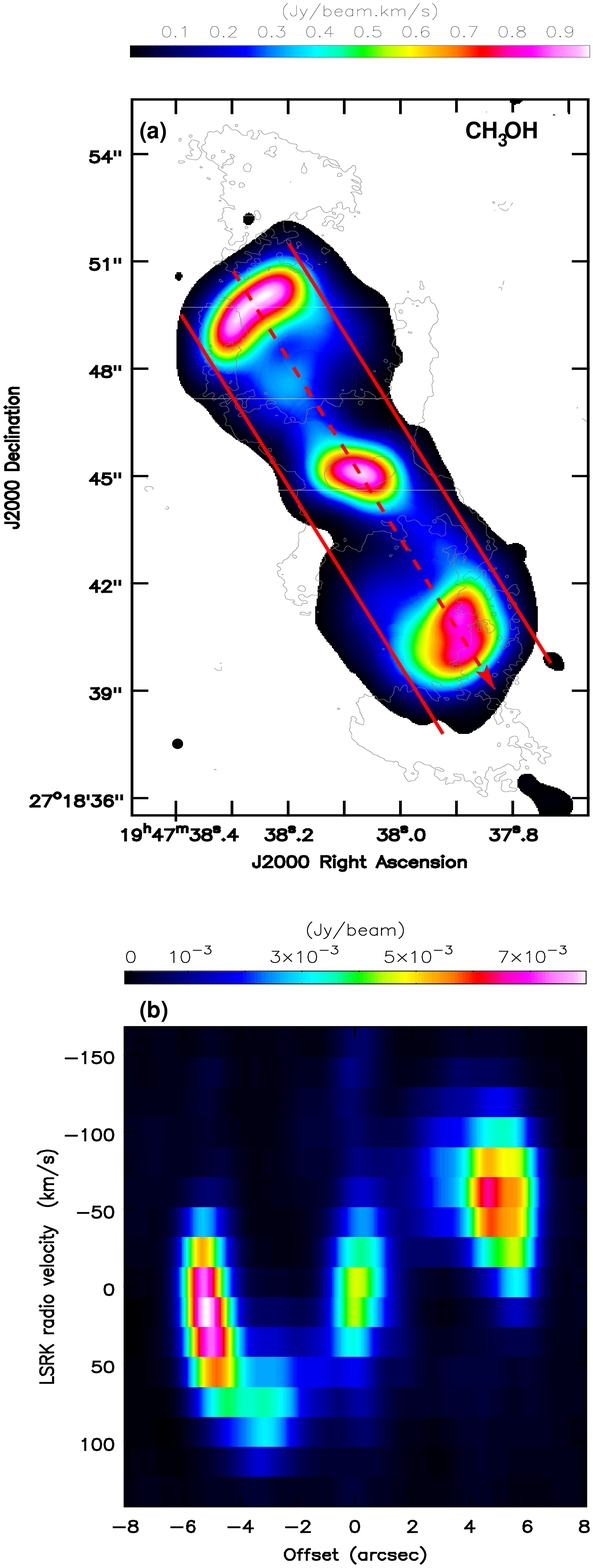}
\includegraphics[width=0.45\textwidth]{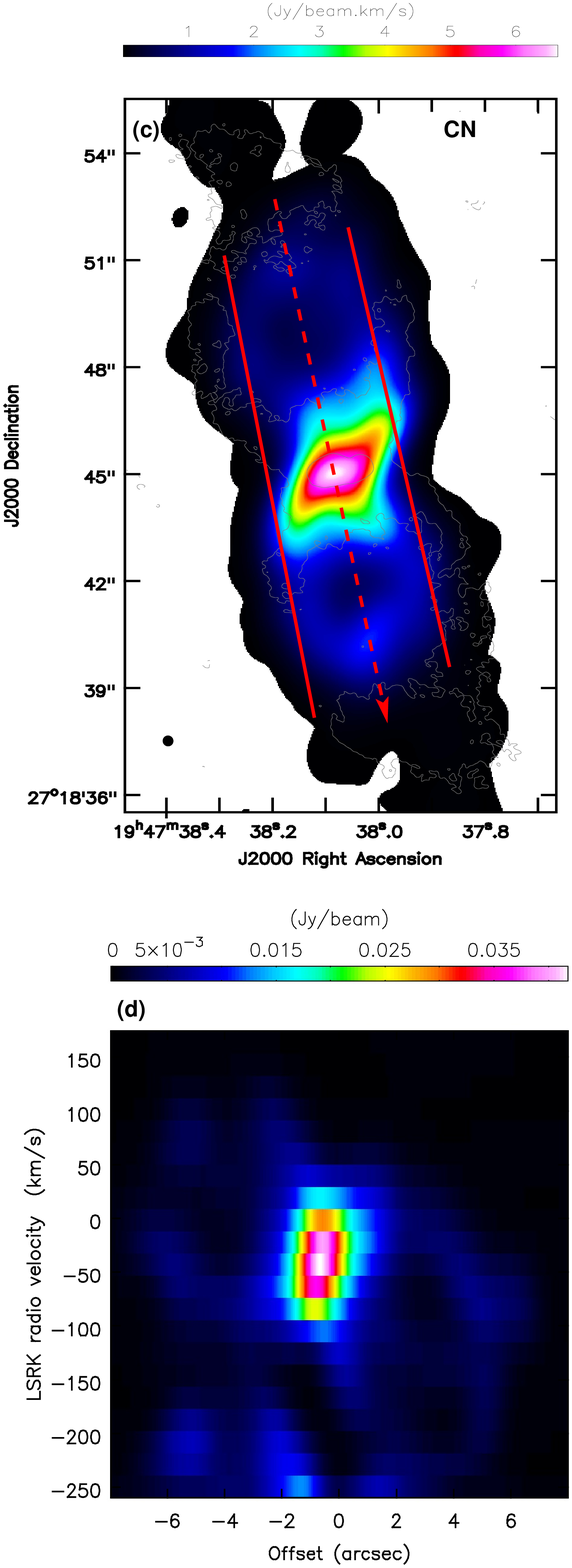}
\caption{(a) Intensity map of 
\textcolor{black}{the 241.787~GHz CH$_3$OH (colour scale, see
  Table~\ref{IDs})}; contours of dust continuum at 0.12 and
  0.48~mJy~beam$^{-1}$ for reference; (b) position--velocity plot
  along the dashed--line with arrow in (a); (c) intensity map of 
  \textcolor{black}{the 226.663~GHz CN line
  (colour scale)}, contours as (a); (d)
  position--velocity plot along the dashed--line with arrow in (b)
  from north to south. For the intensity maps in (a) and (c) the
  spatial resolution is represented by the small black circle to the
  bottom left in each image, where both the spectral line and
  continuum resolutions are over--plotted, but are indistinguishable
  at this scale. 
    \label{maps1}}
\end{figure*}

\begin{figure*}
\includegraphics[width=\textwidth]{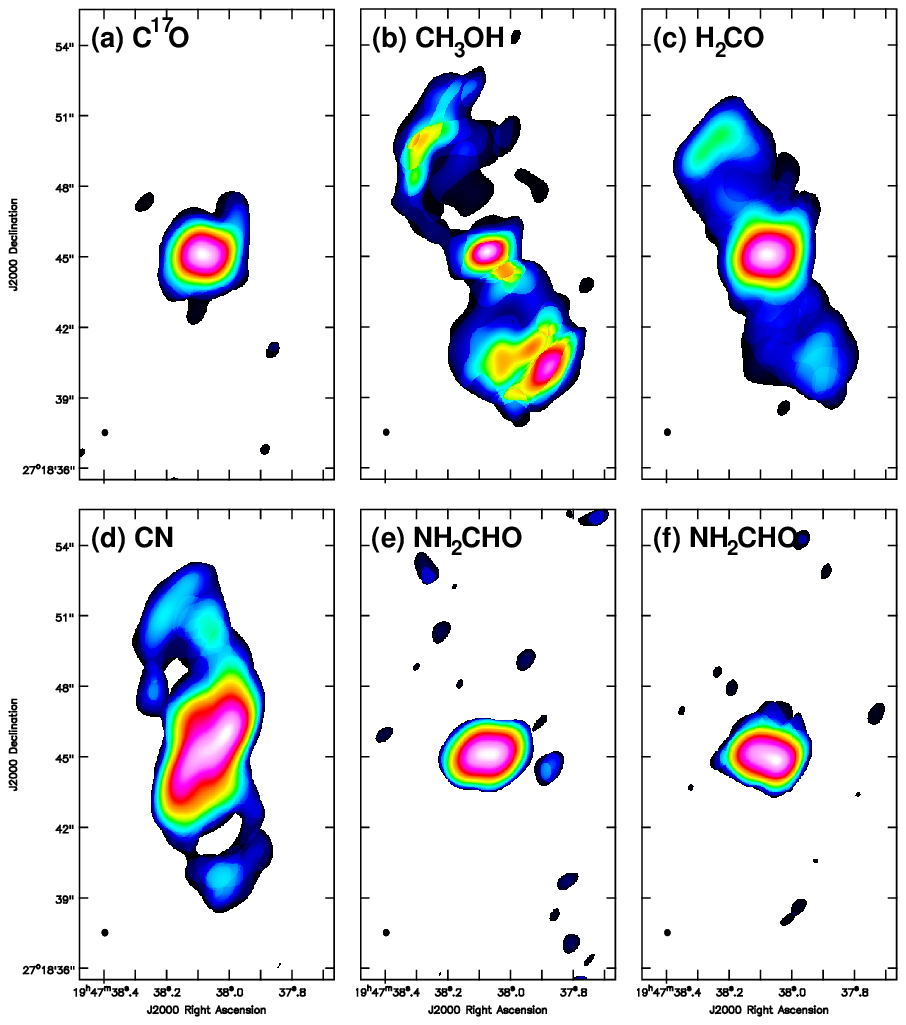}
\caption{Intensity maps of (a) C$^{17}$O; (b) CH$_3$OH; (c) H$_2$CO; (d) CN; 
(e) NH$_2$CHO; (f) NH$_2$CHO. Colour scales are omitted for clarity; 
they differ in
  each case, but cover the full range of the brightness, refer to
  Fig.~\ref{lines} for relative line strengths. The spatial resolution
  is represented by the small black circle to the bottom left in each
  image. Line numbers relate to those in Table~\ref{IDs}.
  \label{maps2}}
\end{figure*}

In order to examine the structures present in the lines in
Table~\ref{IDs} we have taken velocity cuts along the extended
features, which are generally aligned north--south, consistent with
the dust distribution and the CO jet detected by \cite{kaminski-n}. We
have collapsed the data cubes to show the spatial distribution of the
emission.
\textcolor{black}{The CN line at 226.663~GHz and the CH$_3$OH line at 241.787~GHz}
show the clearest extended
structure and are presented here (see Fig.~\ref{maps1}{, where the
  intensity maps show the integrated emission from the spectral data
  cubes, collapsed along the velocity axis}).

\textcolor{black}{The CH$_3$OH line at 241.787~GHz}
shows a triple structure in the collapsed map in
Fig.~\ref{maps1}(a), with bridges of emission between the northern,
central and southern features, together forming an ``S'' shape. {A
  position--velocity plot is included in Fig.~\ref{maps1}(b). This
  runs from north to south along the direction of the dashed
  line, integrating emission across the 99--pixel wide region between
  the parallel solid lines. } This {is} consistent with a jet of gas in a
direction that makes a small angle to the plane of the sky, and either
decelerates or is dramatically redirected as it arrives at the north
and south features, implying that it is encountering a region of
higher density. Comparison with the dust distribution shows that the
{NE--SW extension of the central {CH${_3}$OH} peak is}
consistent with the {direction of the} inner--most dust emission
{(inner--most contour in Fig.~\ref{maps}(a)).} The N and S
features are located with the ends of the dust arcs just before
emission from the clouds begins, i.e. on the border of
these regions closest to the central structure. \cite{kaminski-a}
detect this line, with an identification as CH$_3$OH.

\textcolor{black}{The 226.663~GHz CN line}
shows a structure that is well--aligned with the dusty arcs to
the NW and SE of the central emission. To the NE the line
  emission sits to the west of the adjacent arc, while to SW it sits
  to the east of the adjacent arc (Fig.~\ref{maps1}{(c)}). The
latter two components are better aligned with the jet traced by
CH$_3$OH {(compare (Figs.~\ref{maps1}(a) and (c)). O}verall they
form ``loops'' on the sky that trace the inner boundary of the dust
{structures}. The velocity profile {north to south in the
  direction of the dashed line in Fig~\ref{maps}(c) including the
  99--pixel wide region between the parallel solid lines, is presented
  in Fig.~\ref{maps1}(d). This } also show loops, and {is
  consistent with expanding bubbles. T}aken together this suggests
this gas is arranged in two bubbles expanding at $\sim140$\vunit,
traced north and south of the central feature. \cite{kaminski-a}
identified this line as CN, as well as the adjacent CN line {with
  a peak} that sits just outside SPW1. 
  \textcolor{black}{The 226.338~GHz line} (also CN) shows a
similar, but fainter structure.

Referring to Fig.~\ref{maps2}, 
\textcolor{black}{the lines at 224.273 (C$^{17}$O), 241.621~GHz
 and 242.107~GHz both (NH$_2$CHO)} show only
emission approximately coincident with the {dust emission from the
  compact region} (Fig.~\ref{maps}{(a)}); for 
  \textcolor{black}{the 241.621~GHz line} this
{aligns with} the overall dust distribution, but without the
warped--disc substructure. 
\textcolor{black}{The 225.553~GHz and 225.704~GHz lines}
are more difficult to disentangle, but overall appear to agree with the jet structure traced
by the \textcolor{black}{241.787~GHz CH$_3$OH line;} 
there are elements of the 
\textcolor{black}{226.663~GHz} bubbles present also.

The line--of--sight velocity structures of the bubbles and the jets
are consistent with the gaseous northern features being {angled}
towards the Earth, with the southern features {angled} away,
  consistent with \cite{hajduk-ck}. The orientation of
the dust arcs is harder to constrain. The north--western one
is overlaid with H$\alpha$ emission. {As we argue that} we are
seeing along the limb of a cocoon, the gas in the north angled towards
the Earth is then not affected by local dust extinction, but that
  in the south and to the north--east sits behind the dusty
  hourglass.

\section{Discussion}
\label{sec:discussion}
 
{\cite{hajduk-ck} argue that} the {extended, hour--glass}
H$\alpha$ emitting material was ejected by Nova Vul~1670. We
  suggest that the dust condensed in material that was ejected more
recently, after the central engine had precessed. Further, it seems
clear that mass--loss is ongoing, as {the jets and bubbles traced
  by molecular emission are consistent with outflows} originating in the central engine. The shape of the jet is consistent with
ongoing precession.

{In addition we suggest that the coincidence of the dust arcs,
  which are due to long path lengths through the edges of the dusty
  cocoons, are causing the dark lanes in the H$\alpha$ emission
  through extinction. It is difficult to see why H would be absent
  from only the points on the cocoons that happen to be the edges of
  those structures as they appear on the sky. If this was the case it
  would suggest either these regions have been cleared of Hydrogen
  (and presumably all other) gas or that the conditions do not support
  the excitation of the Balmer lines. We cannot rule this out.

  The molecular gas, the dust and the gas traced by H$\alpha$ in the
  inner region we believe all originate from activity after events of
  the 1670s.  }

The symmetry of the dust emission, the molecular line emission, and
the H$\alpha$ emission all strongly suggest that the central object is
currently, or was in the past, a multiple system \citep[see][for a
review]{orsola}.

\subsection{Stellar mergers}
\label{mergers}

{We discard the possiblity that a stellar merger \citep[see
  e.g.][and references therein]{kaminski-a} between stars at the main
  sequence or giant phases of evolution can account for what is
  observed.

  \cite{hajduk-ck} estimate the current luminosity of the central
  object to be $\sim 1\pm0.5$\Lsun, based on the requirements to
  ionise the radio source they detect. Geometrical effects could push
  this to no more than a few times that value, certainly somewhat less
  than 10\Lsun. As a stellar merger event preserves the luminosity
  available to the pre--merger system in the post--merger system, this
  constrains the nature of the progenitor.

  \cite{bally} model a merger in a multiple stellar system. They
  expect an outburst luminosity $\sim10^5 L_\odot$, and predict
  outflow velocities for typical parameters of a few times
  30\vunit. They also describe a debris disc with a size of around
  1000\au, and cooling timescales for the merged object between 1000
  and 10\,000~yr. Comparing the various characteristics of \ck\ with
  their predictions gives a picture consistent with the parameters
  from the \citeauthor{bally} modelling. However their work relates to
  the formation of high--mass stars through the merger of precursors
  with masses greater than 10~\Msun. Thus the luminosity estimate of
  \cite{hajduk-ck} excludes this as a possible model for \ck; our
  ejecta mass estimate of 0.1\Msun\ or lower is also consistent with a
  low--mass progenitor.}

\cite{metzger} have modelled a group of stars including V838~Mon and
V1309~Sco, with which \ck\ has been compared. They note that a
double--peaked light curve \citep[as was the case for Nova Vul
1670;][]{shara} is a common characteristic of these objects, and
relate the time between light curve peaks $t_{\rm pk}$ (in days) to
the physical characteristics of a stellar merger, which they propose
as the origin of these systems.  Rearranging their Equation~(21), and
substituting for the escape velocity to eliminate the pre--merger binary
orbital radius, one finds the binary mass to be given by
\begin{equation} 
\frac{M_{\rm binary}}{\Msun} \simeq \frac{t_{\rm pk}^2}{4.52\times10^6}\frac{\xi}{\kappa}\frac{v_{\rm ej}}{M_{\rm ej}}
\label{Mbinary}
\end{equation}
where $M_{\rm ej}$ is the ejecta mass in solar masses, $v_{\rm ej}$
the ejecta velocity in \vunit, and $\xi/\kappa\sim2$ for reasonable
values. For \ck\ we have estimated $M_{\rm
  ej}\sim10^{-2}\Msun$. \cite{hajduk-ck2} estimated $M_{\rm
  ej}\sim5\times10^{-2}\Msun$ for a gas--to--dust ratio of 100. Ejecta
velocities in the literature are in the range 60 to 200\vunit; the
time between peaks in 1670 and 1671 gives $t_{\rm pk}\sim270$~d
\citep[estimated from the lightcurve in][by comparing the
dates of the two peaks]{shara}. This gives $40<M_{\rm
  binary}<650$\Msun, a range {excluded by the \cite{hajduk-ck}
  luminosity estimate. We note we have used the lower limit of our
  ejecta mass estimate.}

{We conclude that no stellar merger models for main sequence to giant star
  progenitors can account for the properties of \ck\ either during the
  1670s events or in contemporary observations.}

\subsection{Planet ingestion}
\label{planets}

\cite{retter} invoked a red giant branch (RGB) star ``swallowing'' a
number of planets to account for the 2003 outburst of the transient
V838~Mon, with which \ck\ has been compared \citep{kato2003}.
\cite{soker} outlined a number of fundamental reasons, including the
energy and mass available in a planet, as to why the
\citeauthor{retter} model could not explain V838~Mon. {Other
  modelling of planet ingestion includes \cite{staff2016}, requiring a
  giant star, \cite{bear2011}, with a brown dwarf primary and
  timescales of only a few days for the brightening, much less than
  observed in the 1670s, and \cite{siess} finding a maximum amplitude
  less than 1000, and brightening extending over $\sim10^4$~yr. None
  of these are consistent with the luminosity or timescales present in
  \ck. Thus we exclude planet ingestion by stars in the range from
  brown dwarfs to giants.}

\subsection{White Dwarf merger with a Brown Dwarf}
\label{sec:WD-BD}

{While a stellar merger has been the favoured interpretation for
  \ck\ \citep{kaminski-a} it seems that no primary star with an intact
  envelope prior to 1670 can evolve in a manner consistent with the
  observed characteristics of \ck. We turn to the results in
  Section~\ref{sec:results} to explore an alternative model.

  While the nature of \ck\ prior to 1670 is currently unknown, it is
  indisputable that the morphology and extent of the associated nebula
  is similar to other objects including asymmetric planetary nebulae
  (APN). The shaping of these objects has been progressively
  demonstrated to be due to jet outflows generated following the
  formation of the nebula through the ejection of the stellar
  envelope. \cite{sahai1998} proposed that dust cocoons in CRL~2688
  were formed by jets expanding out into the surrounding
  nebulosity. Subsequent work has found that jets in APN originate
  with a disc around a companion to the central star
  \citep[e.g.][]{Akashi2017a,Akashi2017b,Cardenas2017,Estrella2017,Akashi2018}. The
  observed alignment between the dust cocoons, the jets and the
  bubbles demonstrate that the explanation of APN morphology could
  also apply to \ck. We also believe the innermost structures seen
  with ALMA are consistent with a disc.

  What then is the origin of this disc? We have already excluded the
  sorts of progenitors that would be necessary for a merger involving
  main sequence or giant stars. The common envelope phase that
  precedes PN formation in binary systems would have been visible
  prior to 1670. Thus we need an alternative origin.

  We suggest that all the observed characteristics of \ck\ are
  consistent with a {\em merger between a white dwarf (WD) primary and a
  brown dwarf (BD) secondary}. This would result in the observed
  ``extraordinary'' \citep{kaminski-a} isotopic composition, as the
  ejecta would be the disrupted brown dwarf, subjected to nuclear
  burning at elevated temperatures during the merger event. The
  estimated ejecta mass is also consistent with that of a
  BD. \cite{hajduk-ck} argue that the central object must have an
  effective temperature in the $4-10\times10^4$~K range, consistent
  with a WD that has experienced recent nuclear burning in accreta at
  its surface. They also identify elevated levels of lithium in the
  ejecta, further supporting nucleosynthesis at the point of
  merger. Unlike other merging stellar object models
  \citep[e.g.][]{kaminski-2018}, there is no barrier to prevent the
  products of the nucleosynthesis being ejected from the merger event
  almost immediately, as this takes place on the WD surface. The BD
  material that is not ejected would necessarily form an accretion
  disc due to the orbital angular momentum. This then generates the
  conditions required to support the observed jets at some point after
  the initial ejection events in the 1670s. The jets go on to inflate
  the bubbles and generate the dusty cocoons observed in
  Figs.~\ref{maps1} and \ref{maps} over the subsequent centuries.

}

 \section{Conclusions}
 \label{sec:conclusions}
 
 We have imaged the inner regions of the extended emission associated
 with \ck\ in unprecedented detail. The dust emission {is found to
   trace a warped disc extended east--west about 1400~\au, a pair of
   cocoons extending $\sim5\arcsec$ to the north and south, and two
   additional clouds $2\arcsec$ in extent located $6\arcsec$ to the
   north and south of the central peak, which coincides with
   free--free radio emission detected by \cite{hajduk-ck}.}

 In addition we have imaged extended emission in at least eight lines,
 which are resolved at a modest level of $\sim$22~km~s$^{-1}$. This is
 sufficient to show a jet traced by CH$_3$OH and extended bubbles
 traced by CN, both well--aligned with the dust emission. The
 extension of the inner feature in the central compact structure also
 aligns with the jet axis. {The mapped emission is} consistent
 with a CO jet detected by \cite{kaminski-n}, {as well as
   structures traced in numerous molecules including AlF by
   \cite{kaminski-2018},} and it seems reasonable to assume we have
 resolved it. Some lines detected are consistent with bubbles seen in
 CN, with other molecules tracing the jet. {The S--shape of the
   jet is consistent with precession at an unknown period.} Further
 dynamic modelling of these lines will allow us to better understand
 the interactions between these components; this work will be
 presented elsewhere.

We make comparisons with the predictions of two stellar--merger
models, and note that many of the features seen in Nova Vul~1670 are
consistent with such an origin. {However in general the
  invisibility of the central star prior to 1670 is inconsistent with
  any stellar merger scenario for main sequence or giant stars}, in
line with the conclusions of \cite{smorgasbord}. {We also exclude
  planet ingestion as this cannot explain the duration of the
  brightening events in the 1670s.}

{We suggest instead that Nova~Vul~1670 was due to the merger of a
  white dwarf and a brown dwarf. This is consistent with our mass
  determination for the nebula imaged with ALMA and in H$\alpha$. Jets
  forming dust cocoons and cavities, as seen in \ck, are believed to
  cause similar structures in APN.  We argue that the brown dwarf
  impact generates the unusual abundances and isotopic ratios seen in
  this object via nucleosynthesis, then forms the extended ejecta and
  disc observed with ALMA and that in turn drives the jets shaping the
  inner $\sim$6\arcsec\ north and south of the centre of the jet and
  disc. This would include generating and distributing the $^{26}$Al
  recently imaged by \cite{kaminski-2018}.}

\section*{Acknowledgements}

{We thank the anonymous referee for their helpful feedback which has
improved the paper significantly.}

This paper makes use of the following ALMA data:
ADS/JAO.ALMA\#2016.1.00448.S.  ALMA is a partnership of ESO
(representing its member states), NSF (USA) and NINS (Japan), together
with NRC (Canada), NSC and ASIAA (Taiwan), and KASI (Republic of
Korea), in cooperation with the Republic of Chile.  The Joint ALMA
Observatory is operated by ESO, AUI/NRAO and NAOJ.

Based in part on observations obtained at the Gemini Observatory
(observing program GN-2010A-Q-62), which is operated by the Association
of Universities for Research in Astronomy, Inc., under a cooperative
agreement with the NSF on behalf of the Gemini partnership: the National
Science Foundation (United States), the National Research Council
(Canada), CONICYT (Chile), Ministerio de Ciencia, Tecnolog\'{i}a e
Innovaci\'{o}n Productiva (Argentina), and Minist\'{e}rio da
Ci\^{e}ncia, Tecnologia e Inova\c{c}\~{a}o (Brazil).

RDG was supported by NASA and the United States Air Force.  CEW was
supported in part by NASA \sirtf\ grants to the University of
Minnesota.  SS acknowledges partial support from NASA, NSF and \sirtf\
grants to ASU. MH acknowledges Polish MSHE for funding grants
DIR/WK/2016/2017/05--1 and 220815/E--383/SPUB/2016/2.

\bsp	
\label{lastpage}
\end{document}